# Defect-mediated electron-phonon coupling in halide double perovskite

Aprajita Joshi[1], Sajid Saikia[2], Shalini Badola[1], Angshuman Nag[2], Surajit Saha[1]*

[1]Department of Physics, Indian Institute of Science Education and Research Bhopal, Bhopal 462066, India

[2]Department of Chemistry, Indian Institute of Science Education and Research Pune, Pune 411008, India

*Correspondence: surajit@iiserb.ac.in

**ABSTRACT**

Optically active defects often play a crucial role in governing the light emission as well as the electronic properties of materials. Moreover, defect-mediated states in the mid-gap region can trap electrons, thus opening a path for the recombination of electrons and holes in lower energy states that may require phonons in the process. Considering this, we have probed electron-phonon interaction in halide perovskite systems with the introduction of defects and investigated the thermal effect on this interaction. Here, we report Raman spectroscopy study of the thermal evolution of electron-phonon coupling, which is tunable with the crystal growth conditions, in the halide perovskite systems $Cs_2AgInCl_6$ and $Cs_2NaInCl_6$. The signature of electron-phonon coupling is observed as a Fano anomaly in the lowest frequency phonon mode (51 cm$^{-1}$) which evolves with temperature. In addition, we observe a broad band in the photoluminescence (PL) measurements for the defect-mediated systems, which is otherwise absent in defect-free halide perovskite. The simultaneous observation of the Fano anomaly in the Raman spectrum and the emergence of the PL band suggests the defect-mediated mid-gap states and the consequent existence of electron-phonon coupling in the double perovskite.

The new class of semiconducting material, e.g., lead-based halide perovskite $APbX_3$ where A is the monovalent cation and X is the halide ion, has recently drawn significant attention.[1] Their excellent properties, such as tunable band gap, high photoluminescence (PL) quantum yield, etc., make them a potential candidate for numerous applications such as optoelectronics, photovoltaics, solar cells, light-emitting diodes, etc.[2–6] Alternatively, Pb-free compounds like double perovskites ($A_2MM'X_6$) that replaces $Pb^{2+}$ with MM' where M is the monovalent cation ($Na^+$, $Ag^+$, etc.) and M' is the trivalent cation ($In^{3+}$, $Bi^{3+}$, etc.) are being explored.[7–9] Thus, substituting two elements in place of Pb provides new options for the tunability of its material properties, including different kinds of doping.[10] Even though these materials hold huge potential in optoelectronic applications, they are sensitive to atomic disorders like point defects, which easily form and influence their optical and electronic properties.[11,12] The presence of defects in the crystals often leads to the emergence of mid-gap states (within the band gap) that provide an extra channel for the electrons for recombination and trapping. In addition, these defects in the crystals can also lead to the observation of various fascinating properties, such as single-photon emitters in a wide band gap material like diamond, h-BN, etc.[13,14]

Lead-free halide double perovskite $Cs_2AgInCl_6$,[15] a direct band gap semiconductor (gap ~3.3 eV), stabilizes in the cubic crystal structure.[16] PL studies on this material show a weak PL centered around 2 eV that can be enhanced with doping (say, Na) at the Ag site.[17] The origin of this PL far below the band gap is often explained either by self-trapped excitons or emission due to the presence of intrinsic defects in halide perovskite.[17–19] Zhou *et al.,* in their theoretical and experimental studies, reported a broad PL centered around 1.95 eV, which was attributed to the recombination of electrons from the defect states.[18] Similar observations were also reported by Volonakis *et al.* in their absorption studies, where they observed a broad feature around 585 nm indicating the presence of mid-gap states near about 2 eV mediated by defects in the system.[19] Incorporating defects in the system can also provide an additional channel for the interaction of the phonons and electrons. These interactions can be easily probed by non-destructive techniques like Raman spectroscopy, where the signatures of such interactions are observed as alterations in the characteristics of phonon modes. These observations thus provide evidence for the presence of defects in the material.[20,21] In this letter, using Raman spectroscopy, we report photo-induced electron-phonon coupling (EPC) mediated via defect states in microns-sized crystals (tailored system) of $Cs_2(Ag/Na)InCl_6$. The EPC manifests itself as a Fano anomaly in the line shape of low-frequency Raman phonon mode. Our findings suggest defects in the system as the mediator of the coupling. Notably, the EPC is absent in the defect-free millimeter-sized crystals (controlled system).

The crystals were synthesized using the following two methods. (I) Method A: $Cs_2MInCl_6$ (M = Ag, Na) double perovskite single crystals were synthesized by one-pot hydrothermal inspired by Zhang et al. Anhydrous halide precursor salts, 2 mmol CsCl, 1 mmol MCl, 1 mmol $InCl_3$ and 10 ml of 37% (w/w) concentrated hydrochloric acid (HCl) were loaded in teflon-liner. Salts were dissolved in HCl and then teflon was tightly packed inside an autoclave. It was placed in an oven for 12 hours at 180 °C. Then heating was stopped and the autoclave was allowed to cool naturally to room temperature (27 °C). Nearly 1-2 mm sized big single crystals were obtained which were separated by decanting HCl. Crystals were washed four times with isopropanol and followed by vacuum drying for nearly 3 hours inside a desiccator. Then the dried crystals and their powder were used for further study. Henceforth, we will call these crystals as Crystal A. (II) Method B: The acid precipitation method has been used to synthesize $Cs_2MInCl_6$ (M= Ag, Na) microcrystals as reported by Saikia *et al.*[22] Precursor salts, 0.25 mmol $InCl_3$, 0.25 mmol MCl, are taken in a 15 mL glass vial and 4 mL HCl were added. To dissolve precursor salts, the reaction mixture is heated at 78 °C in a silicone oil bath. Then reaction mixture was continuously stirred for 30 min using a magnetic stirrer. Then 0.5 mmol CsCl (2 times of $InCl_3$ mmol) is added to the above clear reaction mixture which immediately leads to white precipitation. The reaction is kept for another 20 min while continuously stirring. Then the glass vial is allowed to cool to room temperature. Then the precipitate was filtrated and washed 3 times with ethanol. Finally, the sample was dried under an infrared lamp. The powder obtained was stored in a glass vial under ambient conditions for further study. Henceforth, we will call these crystals as Crystal B.

The purity of the samples was checked using the powder x-ray diffraction (XRD) technique using a PANalytical diffractometer with Cu Kα radiation (1.5406 Å), for Crystal A and Bruker D8 Advance X-ray diffraction machine with the radiation source as Cu Kα (1.54 Å), for Crystal B. The refinement was done using high score plus software. The crystal structure was further explored by performing Raman spectroscopy in the backscattering geometry using a Lab RAM-HR Evolution Raman spectrometer coupled with a Peltier-cooled charge-coupled (CCD) detector and to an Nd-YAG 532 nm laser source. The temperature-dependent (80 K- 400 K) Raman spectroscopy measurements were performed using the Linkam heating stage attached to the spectrometer (Model no. HFS600E-PB4). PL measurements were performed using the same setup.

Powder XRD patterns of the quaternary double perovskites $Cs_2MInCl_6$ (B = Ag, Na) millimeter-sized crystals grown by method A (Crystal A) and micron-sized crystals grown by method B (Crystal B), discussed above, are shown in Fig. S1 (supplementary material). Rietveld refinement of the XRD data suggests that the samples crystallize in cubic phase (Fm-3m) comprising of $MCl_6$ and $InCl_6$ octahedra stacked alternate to each other and Cs occupying the voids between the two octahedra, as illustrated in Fig. 1(a). The samples are phase pure. The similar XRD patterns of $Cs_2AgInCl_6$ (Crystal A and Crystal B), shown in Fig. 1(b), indicates that the crystal structure stays identical in both the preparation methods. In contrast, the Raman spectra at room temperature are very different for the two samples, as shown in Fig. 1(c). Crystal A shows four phonon modes (labelled as P1, P2, P3, and P4), whereas Crystal B exhibits three additional modes (A1, A2, and A3).

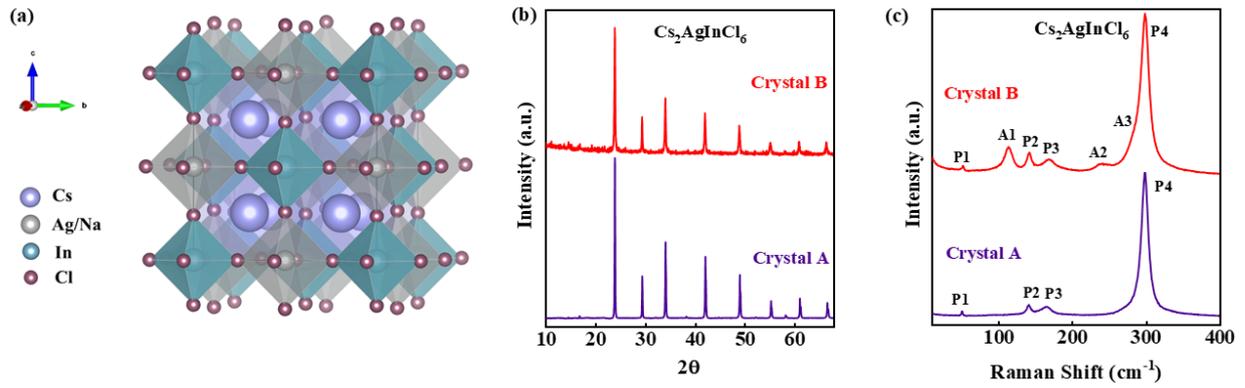

**Fig. 1** (a) The crystal structure of $Cs_2$(Ag/Na)$InCl_6$, (b) comparison of room-temperature XRD, and (c) room-temperature Raman spectra of $Cs_2AgInCl_6$ synthesized by A and B methods.

According to group theory, the irreducible representation of these phonons is: $2F_{2g} + A_g + E_g$.[23] The lowest frequency mode P1 (51 cm$^{-1}$), is assigned as $F_{2g}$ symmetry which describes the translation of Cs atom in the space between the octahedra $AgCl_6$ and $InCl_6$, P2 (113 cm$^{-1}$) is of again $F_{2g}$ symmetry that arises due to the breathing motion of the octahedra, whereas P3 (141 cm$^{-1}$) is associated to $E_g$ symmetry and the high-frequency phonon P4 (288 cm$^{-1}$) is an $A_g$ symmetry mode where $E_g$ and $A_g$ denote the stretching vibrations of the octahedra.[23] Fig. 1(c) (upper panel) shows the Raman spectra of Crystal B. We observe a few marked differences in the Raman response of crystals grown by both methods. Firstly, the lowest frequency mode P1 in the sample shows an asymmetric line shape in Crystal B, while a symmetric lineshape was observed in Crystal A, as shown in Fig. 2(a). Secondly, for crystal B, additional Raman modes were observed (A1, A2, and A3), as mentioned above, which are absent in crystal A, the origin of which will be discussed later.

Typically, the phonons have Lorentzian lineshape,[24] as can be seen for all the phonons except for the P1 mode (Crystal B) which has an asymmetric nature (upper panel) while in the case of Crystal A (bottom panel), symmetric Lorentzian lineshape is observed, as depicted in Fig. 2(a).

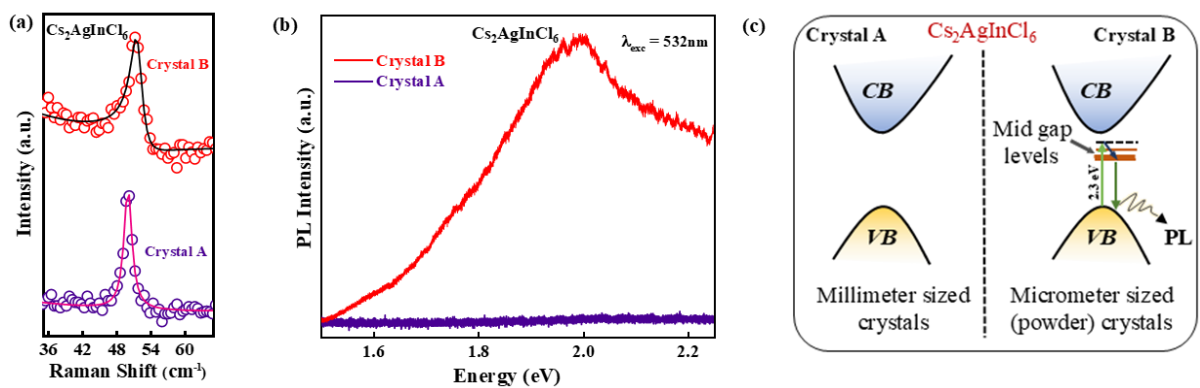

**Fig. 2** (a) Room temperature Raman spectra of $Cs_2AgInCl_6$ (Crystal A and Crystal B), showing a zoomed view of P1 mode. Fano (Crystal B) and Lorentzian (Crystal A) features are depicted by a black and magenta solid line, respectively. (b) Photoluminescence spectra of $Cs_2AgInCl_6$, and (c) schematic representation of $Cs_2AgInCl_6$ defect-free (Crystal A) and defect-rich (Crystal B).

Such an asymmetry results from Fano resonance that occurs due to the interference of a discrete and a continuum state.[24] Generally speaking, a discrete state is due to phonon while the continuum can have electronic or magnetic origin. As these double perovskite systems are non-magnetic, we rule out the possibility of magnetic origin. As the materials under study are wide bandgap semiconductors (3-5 eV), the electronic continuum is not likely to exist. Despite that, we observe the occurrence of Fano asymmetry with a laser energy excitation of 2.3 eV (532 nm), lower than the bandgap suggesting the possibility of the existence of midgap states in the materials. The midgap state provides a level for the photo-induced electronic transition and, thus, the continuum state. Based on the available literature, the origin of these defects may be attributed to Ag/In antisite disorders that give rise to states near about 2 eV.[18] Therefore, the coupling between the P1 phonon (discrete) and defect-mediated electronic continuum gives rise to the Fano interference in Crystal B.

The lineshape for the P1 mode, particularly, is analyzed using Breit Wigner Fano (BWF) model. According to BWF model, the asymmetry in the peak is explained by the intensity equation:[24]

$$I(\omega) = \frac{(\epsilon + q)^2}{1 + \epsilon^2}$$

where $\epsilon$ is given by $\frac{\omega - \omega_0}{\left(\frac{\Gamma}{2}\right)}$ : $\omega_0$ and $\Gamma$ being the frequency and linewidth of the phonon, whereas $\frac{1}{q}$ is the degree of electron-phonon coupling.

We have also performed PL measurements on Crystal A and Crystal B to gain more insight into the defects by exciting the crystals using a laser source of 532 nm (2.33 eV) having energy lower than the bandgap (3-5 eV). In PL measurement, shown in Fig. 2(b), no PL was detected in the case of Crystal A. In contrast, we observed a weak PL feature in Crystal B. The presence of PL peak can be understood as the electronic transition involving midgap states, as shown in Fig. 2(c). The exact nature of the defects are not yet clear, but it might be because of Ag-In antisite, as was suggested by Zhou *et al.*[18] It may be noted that both PL and Fano asymmetry in Raman spectra is seen in the case of Crystal B. However, these are absent in Crystal A, thus suggesting that the electron-phonon coupling in Crystal B is mediated by the defect-induced mid-gap states.

To further verify our hypothesis, we investigated another closely related double perovskite, $Cs_2NaInCl_6$. Similar observations were made for the sample, where a significant distinction was observed in room temperature Raman spectra of Crystal A and Crystal B, as shown in Fig. 3(a). The Raman modes (P1, P2, and P4) were assigned as $F_{2g}$, $E_g$, and $A_g$ (same as for $Cs_2AgInCl_6$). However, the absence of P3 Raman mode in Crystal A could be related to low scattering cross-section. Identical results of phonon asymmetry and PL, as in $Cs_2AgInCl_6$, were also observed in Na-based samples (depicted in Fig. 3(b, c)).

From the observations above, it appears that the PL arising from midgap states and the photo-induced electron-phonon coupling associated with midgap states transitions are strongly related. In order to further elucidate this correlation, we have investigated a series of similar alloyed double perovskite materials $Cs_2Ag_{1-x}Na_xInCl_6$ (x: 0.25, 0.50, and 0.75) and $Cs_2Ag_{0.05}Na_{0.95}BiCl_6$, where we observe a consistent Fano asymmetry and PL, as shown in the SI (supplementary material, Fig. S2, and Fig. S3). Thus, it further suggests the presence of defect-mediated electron-phonon coupling in these materials.

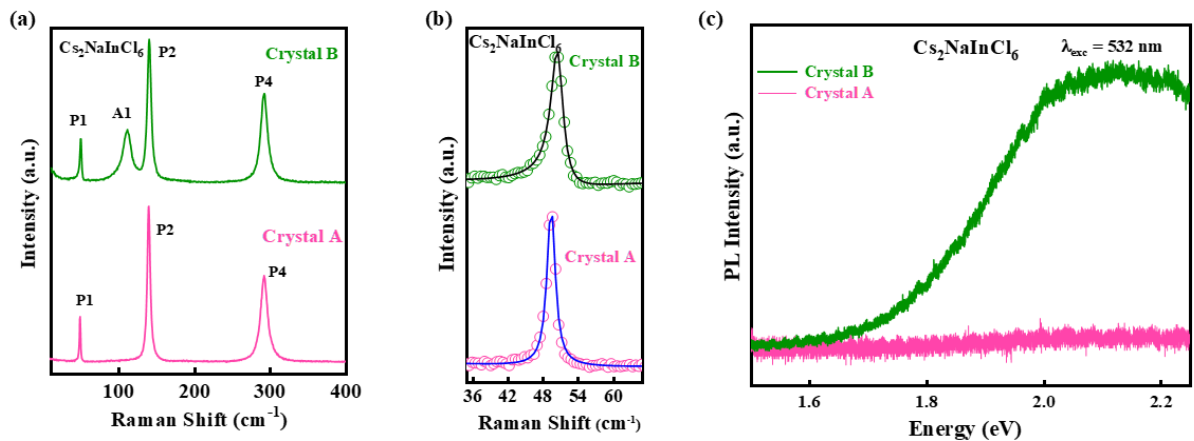

**Fig. 3:** Room temperature Raman spectra of $Cs_2NaInCl_6$ prepared by methods A and B as discussed in the text. (b) A zoomed view of P1 Raman mode, highlighting Fano and Lorentzian fit in black and blue solid lines, respectively. (c) Comparison of PL signal of $Cs_2NaInCl_6$ (Crystal A and Crystal B).

We have further investigated the response of electron-phonon coupling with temperature by carrying out temperature-dependent Raman spectroscopy from 80 to 400 K. The Raman spectra at selective temperatures are shown in supplementary material (Fig. S4 and Fig. S5). It is observed that the electron-phonon coupling shows an appreciable temperature dependence for both the crystals (as shown in Fig. 4).

The effect of temperature on the phonon frequency is explained as:[24]

$$\omega(T) = \omega_0 + \Delta\omega_{qh}(T) + \Delta\omega_{anh}(T) + \Delta\omega_{sp-ph}(T) + \Delta\omega_{el-ph}(T)$$

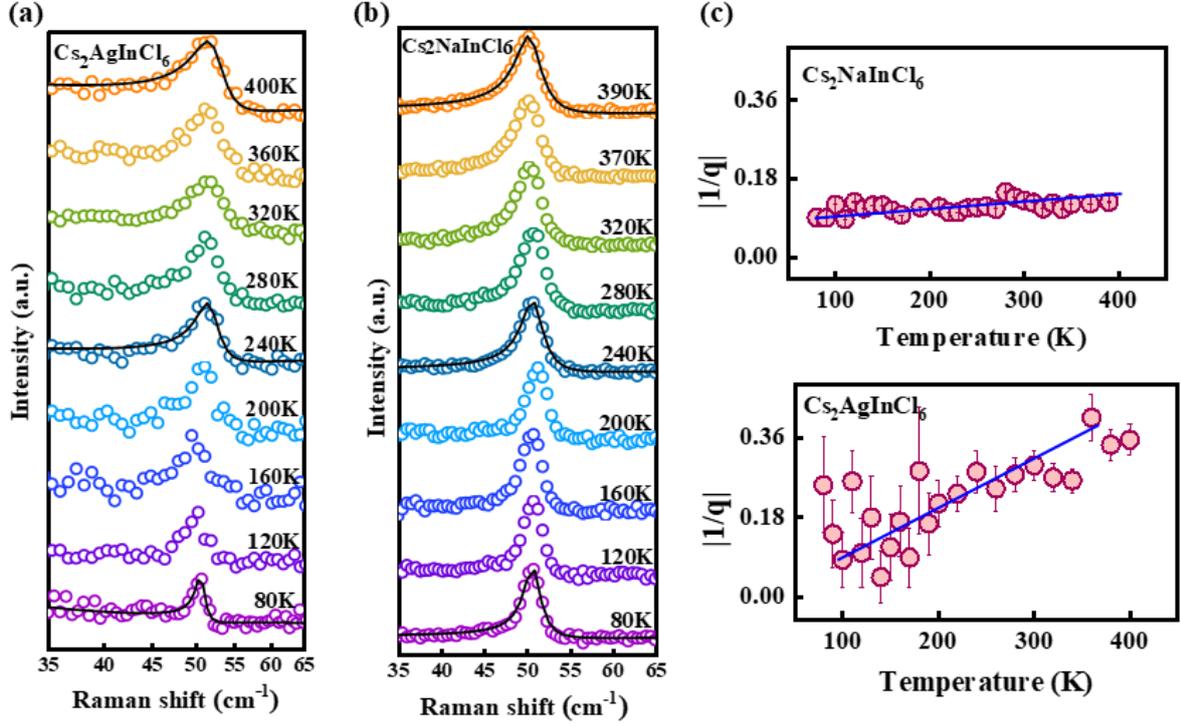

**Fig. 4:** Temperature-dependent Raman of **(a)** $Cs_2AgInCl_6$ (B), **(b)** $Cs_2NaInCl_6$ (B), and **(c)** the comparison of the electron-phonon coupling constant of both the samples (Crystal B).

where the different elements of the equation are explained as $\omega_0$ is the phonon frequency at absolute 0K, $\Delta\omega_{qh}(T)$ indicates the quasi-harmonic contribution in frequency due to the volumetric change in the lattice upon varying temperatures, the consequence of anharmonicity to the crystal is captured by $\Delta\omega_{anh}(T)$. The term $\Delta\omega_{sp-ph}(T)$ is the contribution from spin-phonon coupling which is absent in these non-magnetic systems. The last term of the equation represents the electron-phonon coupling ($\Delta\omega_{el-ph}(T)$).

The cubic anharmonic model, given by Balkanski, suggests that the change in phonon frequency with temperature is:[25]

$$\omega_{ph}(T) = \omega_0 - C\left(1 + \frac{2}{e^{\frac{\hbar\omega_0}{2k_BT}} - 1}\right)$$

where the $\omega_0$ is the phonon frequency at 0 K, C is the constant, and $\hbar$ and $k_B$ are the Planck's and Boltzmann constants, respectively. Any deviation from the anharmonic trend is termed as an anomaly in the phonon mode. Fig. S4 (b) (supplementary material) depicts the phonon behaviour with temperature. Here, the lowest frequency mode, P1, in the case of the $Cs_2AgInCl_6$ (Crystal B), shows the deviation from the cubic anharmonic model throughout the temperature range. Such a deviation may be attributed to the presence of electron-phonon coupling. Besides electron-phonon coupling, the presence of phonon-phonon interaction is also present, however, the former is likely to be dominant in these systems. The anomalous temperature dependence of the frequency of P1 further corroborates our findings on the presence of electron-phonon coupling in the system, suggesting the presence of the defect states. These defects could also be the reason for the observation of additional modes in the Raman spectra of $Cs_2AgInCl_6$ (Crystal B) and $Cs_2NaInCl_6$ (Crystal B) (supplementary material, Fig. S4 and S5).

$Cs_2MInCl_6$ (M: Na, Ag) double perovskites were synthesized by employing two methods to get millimeter-sized Crystal A and micrometer-sized Crystal B. XRD cannot distinguish between Crystal A and Crystal B. However, temperature-dependent Raman spectroscopy shows clear differences between the two kinds of crystals, showing mid-gap defects only in Crystal B. Consequently, Crystal B shows asymmetry in the lowest frequency phonon mode at 51 cm$^{-1}$ because of defect-mediated electron-phonon coupling. Also, the sub-bandgap excitations of defect states lead to PL emission. Clearly, the synthesis method can significantly tailor the defect-mediated electron-phonon coupling in halide double perovskites, and therefore, using appropriate growth techniques could make them potential for optoelectronic applications.

See the supplementary material for XRD, detailed phonon studies, and photoluminescence measurements on $Cs_2AgInCl_6$, $Cs_2NaInCl_6$, and similar alloyed compounds.

## Acknowledgement

S.S. acknowledges Science and Engineering Research Board (SERB) (Grants No. ECR/2016/001376 and No. CRG/2019/002668) and Ministry of Education, India (Grant No. STARS/APR2019/PS/662/FS) for funding. A.J. acknowledges Council of Scientific and Industrial Research (CSIR) for fellowship (09/1020(0179)/2019-EMR-I). A.N. acknowledges Science & Engineering Research Board (SERB, CRG/2022/001199 and SB/SJF/2020-21/02) India. S. Saikia acknowledges Prime Minister's Research Fellowship (PMRF), Ministry of Education, India.

# Supplementary Information

# Defect-mediated electron-phonon coupling in halide double perovskite

Aprajita Joshi[1], Sajid Saikia[2], Shalini Badola[1], Angshuman Nag[2], Surajit Saha[1]*

[1]Department of Physics, Indian Institute of Science Education and Research Bhopal, Bhopal 462066, India

[2]Department of Chemistry, Indian Institute of Science Education and Research Pune, Pune 411008, India

*Correspondence: surajit@iiserb.ac.in

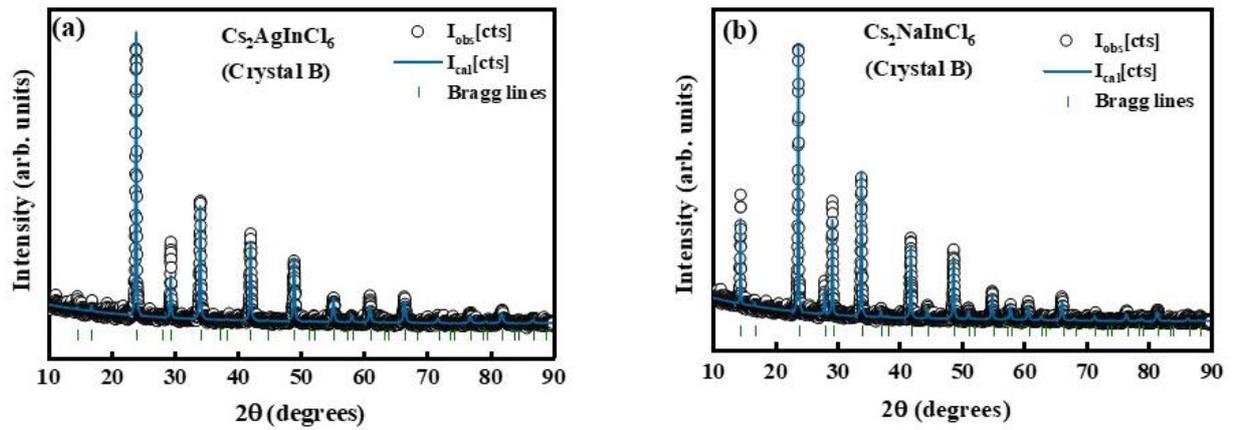

**Fig. S1:** X-ray diffraction pattern at room temperature of **a)** $Cs_2AgInCl_6$ (crystal B) and **b)** $Cs_2NaInCl_6$ (crystal B).

## Observation of Fano asymmetry in similar Halide double perovskite

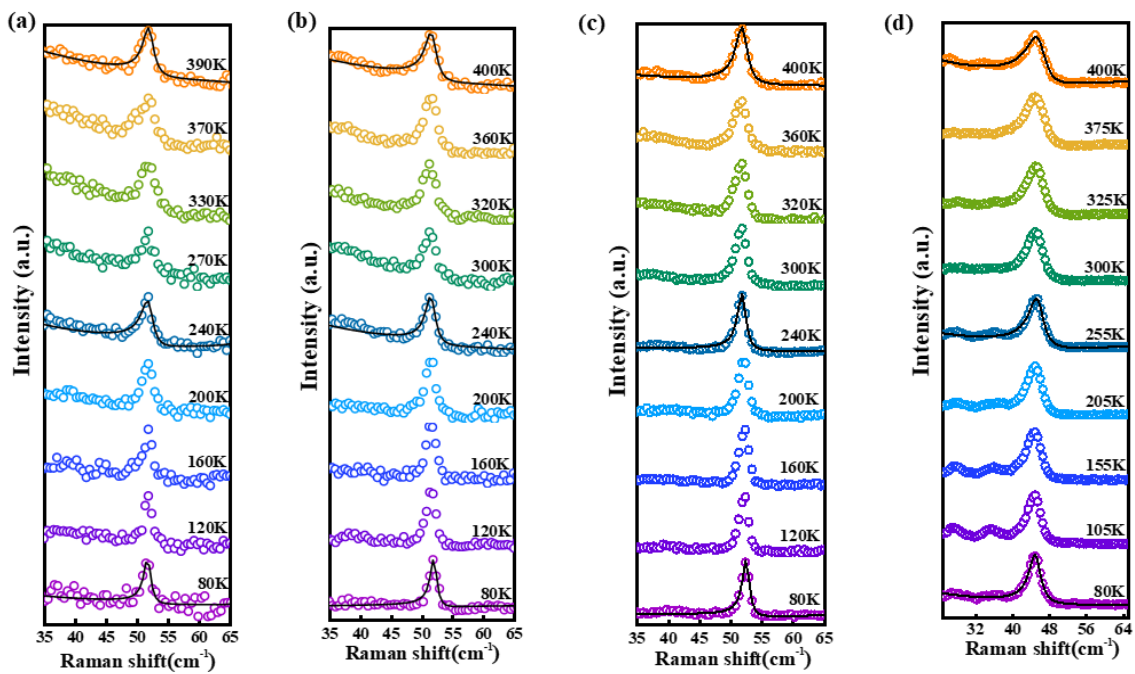

**Fig. S2:** Temperature-dependent Raman spectrum showing evolution of P1 asymmetric phonon mode of (a) $Cs_2Ag_{0.75}Na_{0.25}InCl_6$, (b) $Cs_2Ag_{0.5}Na_{0.5}InCl_6$, (c) $Cs_2Ag_{0.25}Na_{0.75}InCl_6$, and (d) $Cs_2Ag_{0.05}Na_{0.95}BiCl_6$. The black solid line shows the BWF fit.

The presence of Fano asymmetry was also observed in the doped compounds, i.e., $Cs_2Ag_{1-x}Na_xInCl_6$ (x = 0.25, 0.5, and 0.75) and $Cs_2Ag_{0.05}Na_{0.95}BiCl_6$, identical to the crystal B of $Cs_2AgInCl_6$ and $Cs_2NaInCl_6$. We also observe a significant change in the Fano asymmetry with an increase in temperature, as seen in the temperature-dependent Raman spectra of these compounds (illustrated in Fig. S4).

**Photoluminescence in doped halide perovskites**

Fig. S5 depicts the PL spectra of the doped compounds recorded with 532 nm laser excitation. The observation PL signal indicates the existence of defects in the doped systems. The appearance of Fano asymmetry (discussed above) and PL peak suggests the presence of electron-phonon coupling.

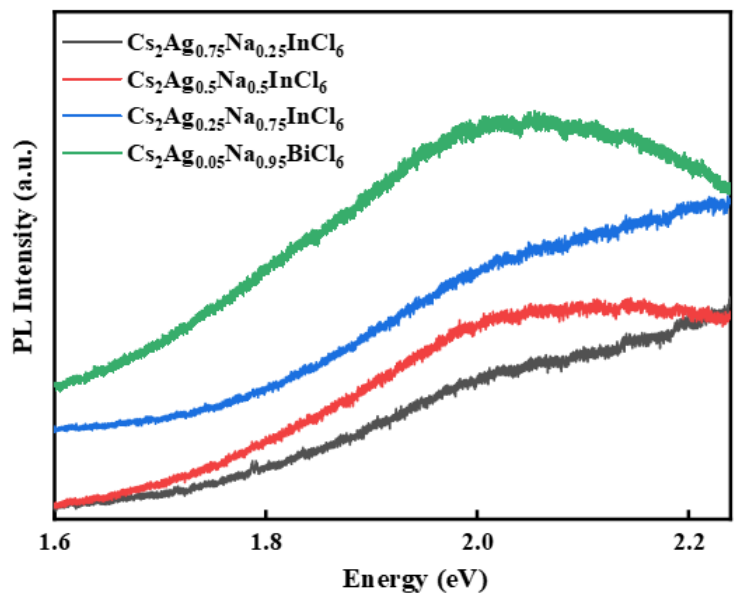

**Fig. S3:** Room temperature PL signal (a) $Cs_2Ag_{0.75}Na_{0.25}InCl_6$, (b) $Cs_2Ag_{0.5}Na_{0.5}InCl_6$, (c) $Cs_2Ag_{0.25}Na_{0.75}InCl_6$, and (d) $Cs_2Ag_{0.05}Na_{0.95}BiCl_6$.

**Temperature variation of phonon frequencies**

Fig. S2(a) shows the temperature-dependent Raman spectra for a few selected temperatures. The temperature variation of the peak position of the phonons follows the anharmonic model as explained in the main text by eq no. 1. Here, the behavior of phonons A1, A2, A3, and P4 is well described by this model. However, the low-frequency mode (P1) depicts an anomalous behavior with an increase in temperature (shown in Fig. S2(b)), which was attributed to the presence of electron-phonon coupling. We also observe an anomalous behavior (for P2 mode) and large Raman shift (for P3 and P4), which could be related to strong anharmonicity. As discussed, Cs atoms occupy large void spaces between the octahedra. It may be noted that the typical dimensions of octahedra along the c axis is 5.04 Å, and along the ab plane is 3.56 x 3.56 Å, whereas those of the void space is 5.25 Å along the c axis and along the ab plane is 7.43 x 7.43 Å. This makes the Cs atoms have large amplitudes of oscillations, thus increasing the anharmonicity. The modes P2, P3, and P4 are associated with the Ag/In-Cl octahedral vibrations, and the Cl ion lie in the vicinity of the Cs atom, which could possibly be the cause of anomalous behavior of P2 and the large shifts in P3 and P4 modes are due to large anharmonicity. This behavior was also consistent for the defect-free sample (crystal A), thus suggesting that strong anharmonicity is inherently present in these systems. Similar behavior of Raman modes was observed for $Cs_2NaInCl_6$ (crystal A and crystal B), depicted in Fig. S3.

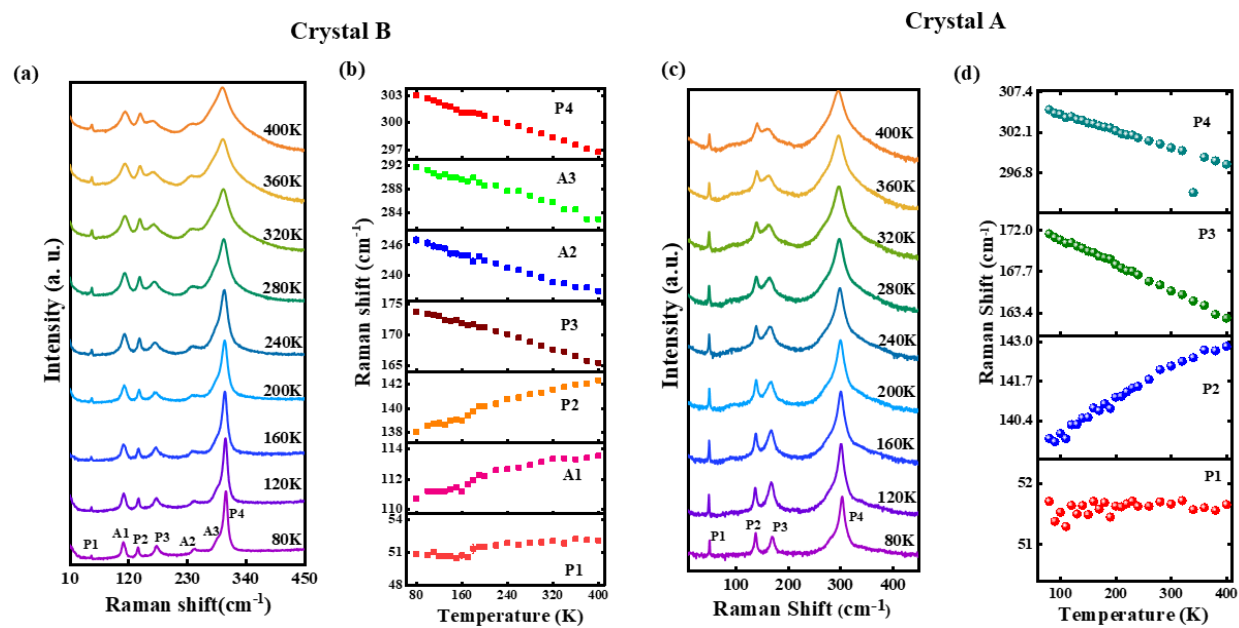

**Fig. S4: a)** Temperature-dependent Raman spectra and **b)** temperature dependence of phonon frequencies of $Cs_2AgInCl_6$ (crystal B). **(c)** Temperature-dependent Raman spectra of and **d)** temperature dependence of phonon frequencies of $Cs_2AgInCl_6$ (crystal A).

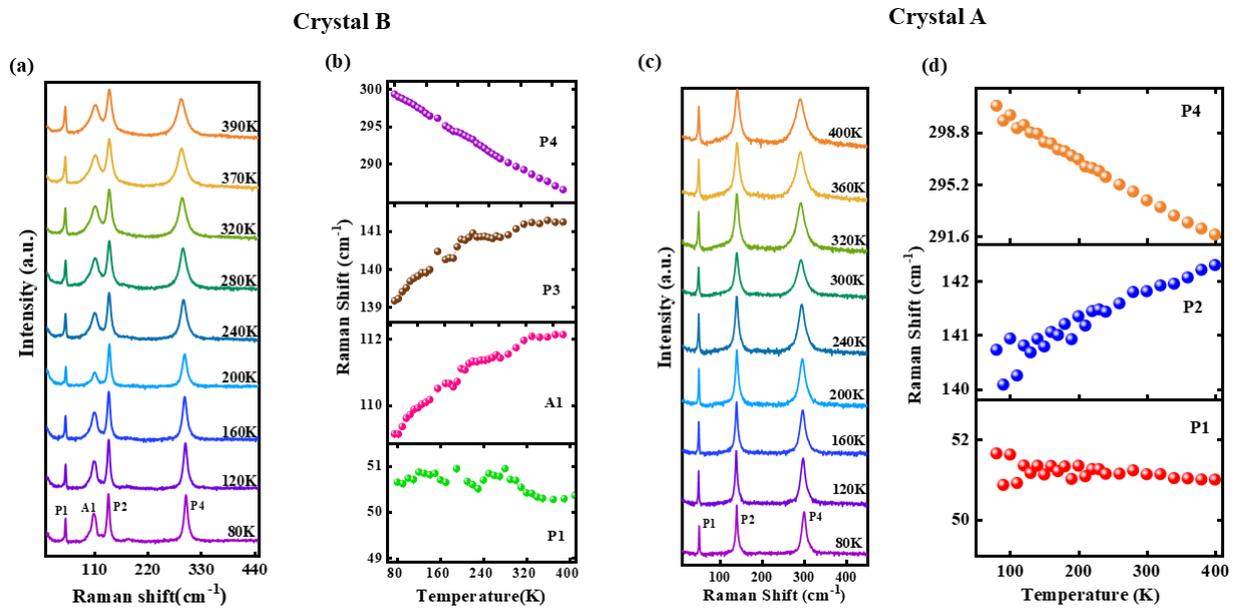

**Fig. S5: a**) Temperature-dependent Raman spectra and **b**) temperature dependence of phonon frequencies of $Cs_2NaInCl_6$ (crystal B). **(c)** Temperature-dependent Raman spectra of and **d)** temperature dependence of phonon frequencies of $Cs_2NaInCl_6$ (crystal A).